\documentclass[10pt,aps,prl,twocolumn,showpacs,floatfix]{revtex4-1}

\usepackage{graphicx}
\usepackage{amsmath}
\usepackage{color}
\usepackage{epsfig}
\usepackage{epstopdf}

\begin{document}

\title{Observations of Dispersion Cancellation of Entangled Photon Pairs}
\author{Kevin A. O'Donnell} 
\affiliation{Divisi\'on de F\'isica Aplicada, Centro de Investigaci\'on Cient\'ifica y de Educaci\'on Superior de Ensenada, Carretera Ensenada-Tijuana N\'umero 3918, 22860 Ensenada, Baja California, M\'exico}

\date{\today}
\begin{abstract}
An experimental study of the dispersion cancellation occurring in frequency-entangled photon pairs is presented.  The approach uses time-resolved up conversion of the pairs, which has temporal resolution at the fs level, and group-delay dispersion sensitivity of $\approx \! 20 \, \mathrm{fs}^2$ under experimental conditions.  The cancellation is demonstrated with dispersion stronger than $\pm 10^3 \, \mathrm{fs}^2$ in the signal $(-)$ and idler $(+)$ modes.  The observations represent the generation, compression, and characterization of ultrashort biphotons with correlation width as small as 6.8 times the degenerate optical period. 

\end{abstract}
\maketitle

The temporal dispersion of classical and quantum states of light has been of wide interest.  In the quantum case, two remarkable types of dispersion cancellation have been noted for pairs of frequency-entangled photons.  First, in the Hong-Ou-Mandel two-photon interferometer \cite{HOMI87}, it has been observed that the interferogram is independent of any external group delay dispersion (GDD) experienced  by the incident two-photon state \cite{Stein92}.  The effect was physically explained in Ref.~\onlinecite{Stein92} through the frequency-dependence of the Feynman paths occurring within the interferometer.  A second, distinct type of dispersion cancellation has been discussed theoretically by Franson \cite{Franson92}.  Here, if one photon of the state is given some amount of GDD, and the second is given an equal but opposite amount, the effects cancel and the two-photon state \textit{itself} is left unchanged.

In subsequent work, the first type of dispersion cancellation has been observed, even, for example, with an interferogram as narrow as $7.2 \, \mathrm{fs}$  \cite{Nasr08}.  On the other hand, the second type of dispersion cancellation has not been observed with comparable resolution.  At the least, the GDD insensitivity of the two-photon interferometer makes it of no use in observing the Franson effect.  Instead, the broadening of Glauber's $G^{(2)}(\tau)$ correlation function \cite{Glauber63} may be used to observe dispersion of photon pairs \cite{Valencia02}.  This approach has been employed in observing Franson's dispersion cancellation with a single optical fiber as the dispersive element \cite{Brendel98} and in an experiment where partial dispersion cancellation was observed  \cite{Baek09}.  

References~\onlinecite{Valencia02}-\onlinecite{Baek09} measure $G^{(2)}(\tau)$ from coincidence rates of detected photon pairs, using detectors with limited temporal resolution.  These resolutions range from 0.2 to 0.8 ns, which implies that quite strong dispersion (in optical fibers of 0.5-6.5 km length) is necessary to produce resolvable effects.  These efforts are not fully satisfying, because dispersive effects are better-studied on a fs-level.  However, detector resolution is a long-standing problem and it seems difficult to make a significant improvement.

Thus, in the following work, a different approach is taken and Franson's dispersion cancellation is studied with $\approx \! 5$ orders of magnitude better time resolution.  The method uses upconversion of entangled photon pairs \cite{Dayan05}, which has become feasible for crystals with strong nonlinearities.   The upconversion rate is monitored as a function of delay $\tau$ between a photon and its pair member \cite{OD09}, with demonstrated dispersion sensitivity equivalent to a path of a few mm in optical glass.  This technique is analogous to autocorrelation methods of classical pulse characterization \cite{Diels06} where fs-level resolution is common, since $\tau$ is determined by mirror position rather than electronic delay.  In one sense, the work here parallels that of Ref.~\onlinecite{Stein92}, but with interference replaced by upconversion.  In another sense, the efforts can be considered a response to a proposal by Harris \cite{Harris07}, although this will be discussed later.  The approach is to produce a given two-photon state and then disperse the two modes as desired, in contrast to the less direct procedures of Refs.~\onlinecite{Brendel98}-\onlinecite{Baek09}.

Consider the two-photon state $| \Psi_{dc} \rangle$ produced by spontaneous parametric downconversion (SPDC) with a monochromatic pump of frequency $\omega_p$
\begin{eqnarray}
| \Psi_{dc} \rangle  \propto  \int d \omega_s  \,  \Phi(\omega_s, \omega_p - \omega_s) \, & e^{i[ \phi_s(\omega_s) + \phi_i(\omega_p - \omega_s)]} 
 \nonumber \\ & \times \,  |\omega_s\rangle_s \,  |\omega_p - \omega_s \rangle_i \, , \label{SPDCState}
\end{eqnarray}
where $\Phi$ is the phase-matching function of the downconversion crystal, and $s$ and $i$ denote, respectively, the signal (frequency $\omega_s$) and idler (frequency $\omega_p-\omega_s$) modes.  The modal phases $\phi_{s,i}$ are those accumulated within the nonlinear crystal producing the state, and in subsequent propagation.  Franson's dispersion cancellation can be considered the preservation of simultaneity of pairs \cite{Franson92} although, more fundamentally, it is the GDD-invariance of this quantum state \textit{itself} \cite{Franson07}.  By changing the integration variable to $\Delta\omega = \omega_s - \omega_d$ with $\omega_d \! \equiv \! \omega_p/2$, the exponential in Eq.~(\ref{SPDCState}) has phase
\begin{equation}
 \phi_s(\omega_d + \Delta\omega) + \phi_i(\omega_d - \Delta\omega)  \! =  \!
\sum\limits_{n = 0}^\infty  \frac{ [\phi_s^{(n)}  \! + (-1)^n \phi_i^{(n)}] \Delta\omega^n} {n!}	\label{SPDCPhase}
\end{equation}
where $\phi_{s,i}^{(n)}$ denotes $\frac {\partial^n \phi_{s,i}(\omega)} {\partial  \omega^n}  \big|_{\omega_d} $.  
The $n \! = \! 2$  term of Eq.~(\ref{SPDCPhase}) represents GDD and often dominates temporal widths; it is clear that the term vanishes if $\phi_s^{(2)} \! = \! - \phi_i^{(2)}$, which is the cancellation condition of Franson \cite{Franson92, Franson07}.  Since this cancellation occurs in the state \textit{itself}, the effect is obviously nonlocal when the signal and idler are spatially separated.  However, the technique employed here requires pairs to be brought back together, and thus represents a \textit{local} observation of this nonlocal effect. 

Now, consider the pairwise upconversion of the state of Eq.~(\ref{SPDCState}) in a second crystal, with a signal/idler delay $\tau$ introduced.  Neglecting the transverse wavevector degree of freedom, the upconverted state is given by \cite{OD09}
\begin{eqnarray}
| \Psi_{uc} \rangle  \propto  \int d \omega_s  \,  |\Phi(\omega_s, \omega_p - \omega_s)|&^2 \,
 e^{i[ \phi_s(\omega_s) + \phi_i(\omega_p - \omega_s)]}
 \nonumber \\ & \times \,  e^{i \omega_s \tau} \,  | \, \omega_p \rangle \, , \label{UCState}
\end{eqnarray}
where it is assumed that the second crystal has properties identical to the first, which introduces a factor of $\Phi^*$, so $|\Phi|^2$ now appears in the integrand.  It is seen that $| \Psi_{uc} \rangle$ has returned to $\omega_p$ but has retained the phase terms of $| \Psi_{dc} \rangle$ in Eq.~(\ref{SPDCState}).  Thus any dispersion cancellation in $| \Psi_{dc} \rangle$ will exhibit direct consequences in $| \Psi_{uc} \rangle$.  Moreover, these effects can be readily observed, since the upconversion rate $R(\tau)$ is proportion to the squared modulus of the integral of Eq.~(\ref{UCState}).

Here, the experimental principles are straightforward: to create photon pairs in a given state, introduce oppositely-signed GDD in the signal and idler modes, and recombine the pairs.  The experiment is shown in Fig.~1 and requires a two-sided extension of a previous approach \cite{OD09}.  The pump laser (power 1 W, wavelength 532 nm, single-frequency) was focused to a $45 \, \mu \mathrm{m}$ waist in an MgO-doped, periodically-poled, lithium niobate crystal of length 5 mm.  The crystal was temperature-controlled ($\approx \! 50^\circ \, \mathrm{C}$) to phase match to co-polarized, frequency-degenerate, axial photon pairs.  The SPDC emission fell within a cone of $\approx 3^\circ$ half-angle, and an iris and 75 mm focal length lens collimated the central $ 2.3^\circ$ (6.0 mm diameter) region.  The bandwidth was measured as $117 \, \mathrm{nm}$, centered on degenerate wavelength $1064 \, \mathrm{nm}$.

\begin{figure}
\includegraphics[width=0.48 \textwidth]{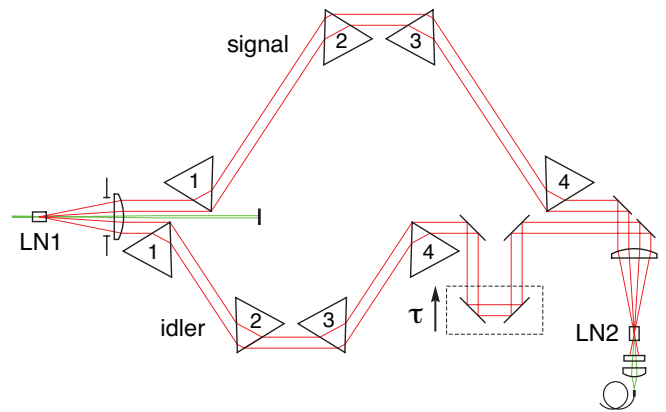}
\caption{ Experimental geometry (not to scale).  SPDC from lithium niobate crystal LN1 is collimated by a lens; the upper beam (signal side) is sent through a prism compressor, and the lower beam (idler side) is sent through a shorter prism compressor and a four-mirror adjustable-delay path.  Positive $\tau$ shortens the path and thus advances the idler or, equivalently, delays the signal photon.  The paths are reunited, reflected downward, and focused to upconvert in crystal LN2.}
\end{figure}

In the collimated beam, transverse momentum conservation implies that a given photon will have its pair member directly opposite beam center.  As shown in Fig.~1, pair members were thus separated by inserting the tips of two prisms above and below beam center, directing the upper beam (signal photons, say) upward and the lower beam (idlers) downward, while the pump beam passed both tips and was absorbed.  Each prism was the first of four prisms ($60.0^\circ$ apex, $30 \, \mathrm{mm}$ sides, Schott SF10 glass) set in the symmetrical, minimum deviation arrangement used to adjust GDD \cite{Diels06}.   The tip-to-tip spacing between the first (or second) pair of prisms was $500 \, \mathrm{mm}$ and $352 \, \mathrm{mm}$ in, respectively, the signal and idler paths.  In addition, the idler photons reflected from four silver mirrors, with the middle two mounted on a translation stage to control $\tau$.  Both branches then returned to their initial interspacing and reflected from two separate silver mirrors, whose adjustment corrected small relative angular misalignments.  This light was focused by a second lens into a second crystal, with both identical to the first.  The upconverted light passed through a BG39-glass filter, and was coupled to a multimode fiber leading to a SPCM-AQR-13-FC photon-counting module.

All prisms were mounted on stages providing translation perpendicular to the prism base, which allows GDD to be changed by known amounts. The principle is simple: a four-prism sequence is often employed to produce negative net GDD, but translating a prism into the beam increases its internal glass path, and the net GDD increases proportionally \cite{Diels06}.  For the system of Fig.~1, direct ray-tracing has shown that the GDD increases by $105 \, \mathrm{fs}^2$ per mm of increased glass path in any prism.

More elaborate procedures are employed here to produce essentially \textit{known values} of $\phi_{s,i}^{(2)} \, $.  In particular, when the signal and idler follow the same path and so suffer the same dispersion, the upconversion rate has been observed to maximize when $\phi_s^{(2)}$  and $\phi_i^{(2)}$ are nearly zero, for the path at $\omega_d$ from the center of the first crystal to the center of the second \cite{OD09}.  In calculations, the case of Ref.~10 (using different prisms) was optimized for $\phi_s^{(2)} \! = \! \phi_i^{(2)} \! = \! 28 \, \mathrm{fs}^2$, which compensated higher-order dispersion.  These values are case-dependent but small, and it will simply be taken here that maximized upconversion implies \textit{zero} $\phi_{s,i}^{(2)} \, $.  The consequences of this simplification are minor, and only imply that values of $\phi_{s,i}^{(2)}$ quoted hereafter should have small constants added.

The notation adopted is that, for example, $\mathrm{P}3_{i}$ refers to the third idler prism in Fig.~1.  To apply the principles discussed, prisms $\mathrm{P}1_{s}$ and $\mathrm{P}4_{s}$ were translated into the beam and $\mathrm{P}1_{i}$ was withdrawn, until \textit{all} SPDC light followed the upper path of Fig.~1. The upconversion rate was then optimized by adjusting insertions of $\mathrm{P}1_{s}$-$\mathrm{P}4_{s}$, so $\phi_{s,i}^{(2)} \! = \! 0$ in this configuration.  $\mathrm{P}1_{s}$ and $\mathrm{P}4_{s}$ were then withdrawn to their original positions, and $\mathrm{P}2_{s}$ and $\mathrm{P}3_{s}$ were inserted by opposite amounts, thus maintaining $\phi_s^{(2)} \! = \! 0$ since the total glass path was unchanged.  $\mathrm{P}1_{i}$ was reinserted so that the idler photons followed the lower path, and $\tau$ was adjusted until upconversion was observed.  The insertions of $\mathrm{P}2_{i}$-$\mathrm{P}4_{i}$ were then set for optimal rate and thus $\phi_{i}^{(2)} \! = \! 0$ (here $\tau$ also had to be varied). Throughout, broadening effects of prism glass path changes of $0.2 \, \mathrm{mm}$ from optimum were observable, which correspond to GDD changes of only $21 \, \mathrm{fs}^2$. Again, this is small compared with GDD introduced later.

Thus, in this optimal state, $\phi_{s,i}^{(2)} \! = \! 0$ and $\tau \! = \! 0$ with reasonable accuracy.  To take data, only three adjustments were required to control $\phi_s^{(2)}$, $\phi_i^{(2)}$, and $\tau$.  Prisms $\mathrm{P}3_{s}$ and $\mathrm{P}3_{i}$ were mounted on computer-controlled stages of step size, respectively, $0.01 \, \mu \mathrm{m}$ and $0.10 \, \mu \mathrm{m}$.  Thus withdrawing $\mathrm{P}3_{s}$ produced negative $\phi_s^{(2)}$, and inserting $\mathrm{P}3_{i}$ produced positive $\phi_i^{(2)}$, with values known to high accuracy.  For simplicity, all prism translation was done so as to change glass path in multiples of $3.50000 \, \mathrm{mm}$ (which, from geometrical optics, required translator changes of $2.08763 \, \mathrm{mm}$ for prism parameters quoted), thus producing steps in $\phi_{s,i}^{(2)}$ of $\Delta \! = \! 367 \, \mathrm{fs}^2$.  Glass path changes must be compensated in $\tau$, and the mirror pair adjusting $\tau$ was mounted on a third computer-driven stage with $0.01 \, \mu \mathrm{m}$ step size, corresponding to $\approx \! 0.07 \, \mathrm{fs}$ steps in $\tau$.  

The signal or idler side power was $32\, \mathrm{nW}$, implying a photon flux of $1.7 \! \times \! 10^{11} \, \mathrm{sec}^{-1}$ in each.  The number of photons per spectral mode \cite{Dayan05} follows as $n \! = \! 0.0055$, consistent with the $n \! \ll \! 1$ isolated-pair limit implicit in Eq.~(\ref{UCState}).  This conclusion is supported by the upconversion rate's linear dependence on pump power, showing a log-log slope of $0.97$ at $1 \, \mathrm{W}$ pump.  Further, the upconversion rate of unentangled photons is not only of order $n^2$, but is smaller still due to a bandwidth factor \cite{Dayan05}; this is consistent with data since $R(\tau)$ will be seen to decay to essentially zero.  Data shown are average rates over $ 6 \, \mathrm{s}$, with statistical errors from $\pm 17 \, \mathrm{s}^{-1}$ in highest signals to $\pm 6 \, \mathrm{s}^{-1}$ in baselines.  The background rate was measured before and after each scan in $\tau$,  with the average subtracted from data.  This was achieved by blocking the path where the signal/idler sides reunite in Fig.~1 and determining the rate ($\approx \! 165 \, \mathrm{s}^{-1}$, nearly the intrinsic detector dark rate).  Then, the beam-block was replaced by a BG39 filter with negligible infrared but high green transmission.  The increase in rate determined the level of stray pump light ($\approx \! 10 \, \mathrm{s}^{-1}$), which was added to the beam-block rate to obtain the background.

Figure~2 shows $R(\tau)$ in cases without dispersion cancellation.  For $\phi_{s,i}^{(2)} \! = \! 0$, $R(\tau)$ has a peak of nearly $1500 \, \mathrm{s}^{-1}$ at $\tau \! = \! 0$ and presents secondary maxima near $\pm 41 \, \mathrm{fs}$, with width and shape similar to previous reports \cite{OD09}.  In the other cases shown, GDD is introduced in steps of $\pm \Delta$ into \textit{either} the signal $(-)$ \textit{or} idler $(+)$ mode, so no cancellation is possible.  The peak first falls to about half its original height (cases (b) and (e)), and continues to fall and broaden as the magnitude of GDD is increased.  In the two cases with strongest GDD effects, the maxima of $R(\tau)$ are $15 \%$ or less of that with $\phi_{s,i}^{(2)} \! = \! 0$, and minima fall near curve center.  Thus Fig.~2 serves to demonstrate that, without cancellation,  the effects of GDD are clear in $R(\tau)$, in distinct contrast to what follows.

\begin{figure}
\includegraphics[width=0.48 \textwidth]{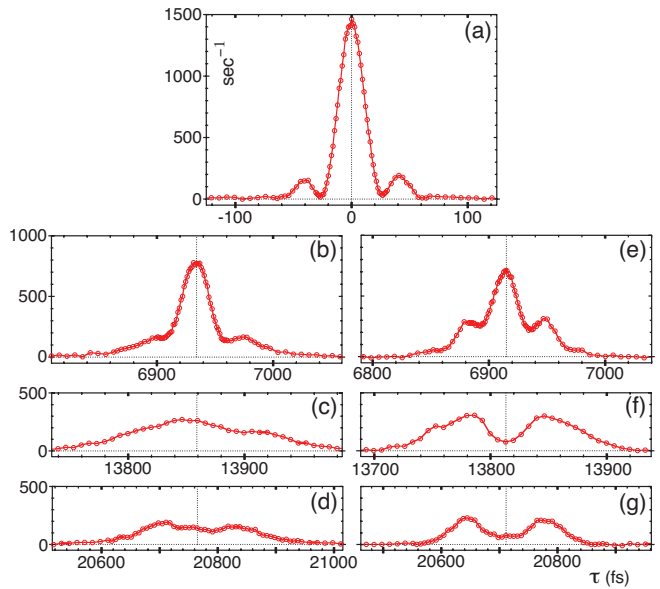}
\caption{ Cases showing dispersion sensitivity in $R(\tau)$.  Results are for  $\{\phi_s^{(2)}, \phi_i^{(2)}\}$ equal to  $\{0,0\}$ (a), $\{-\Delta,0\}$ (b), $\{-2\Delta,0\}$ (c), $\{-3\Delta,0\}$ (d), $\{0,\Delta\}$ (e), $\{0,2\Delta\}$ (f), and $\{0,3\Delta\}$ (g), with $\Delta \! = \! 367 \, \mathrm{fs}^2$.  Dashed vertical lines denote centroids; note $2\times$ $\tau$-width change in (d) and (g).}
\end{figure}

The principal results are shown in Fig.~3, which demonstrate dispersion cancellation in $R(\tau)$.  The values of $\phi_{s,i}^{(2)}$ from Fig.~2 are again used, but the cases of Fig.~3 pair them in the manner expected to produce cancellation.   Throughout all cases, even with $\phi_{s,i}^{(2)}$ at levels producing strong effects in Fig.~2, there is little variation of curve shape in Fig.~3. The mean peak width (full-width at half-maximum) is $24.3 \, \mathrm{fs}$, with all cases within a fs of the mean.  A modest reduction ($11 \%$) of peak height occurs between cases (a) and (d) of Fig.~3; signal levels could be recovered with a small tilt of one of the final mirrors in Fig.~1, probably due to inadvertent tilt introduced by the $\tau$-stage for the long path compensation ($\approx \! 12.4 \, \mathrm{mm}$ total) required.  However, to maintain $\tau$-fidelity, the experiment was left untouched throughout all data of Figs.~2-3.  In summary, the nearly invariant curves of Fig.~3 are a striking result, and represent a direct manifestation of dispersion cancellation within the two-photon state itself.

\begin{figure}
\includegraphics[width=0.48 \textwidth]{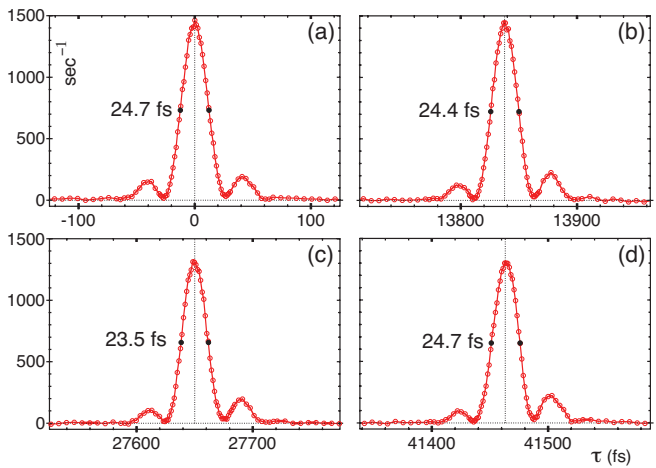}
\caption{ Demonstration of dispersion cancellation in $R(\tau)$.  Results are for  $\{\phi_s^{(2)}, \phi_i^{(2)}\}$ equal to  $\{0,0\}$ (a), $\{-\Delta,\Delta\}$ (b), $\{-2\Delta,2\Delta\}$ (c), and $\{-3\Delta,3\Delta\}$ (d), with $\Delta \! = \! 367 \, \mathrm{fs}^2$.  Dashed vertical lines denote peak centroids, filled circles denote half-height points, and widths are indicated.}
\end{figure}

As a secondary observation, which was first achieved with the two-photon interferometer \cite{Stein92}, it is possible to determine the single-photon speed in glass from the peak position in $R(\tau)$.  It can be shown from geometrical optics that, when a prism is inserted so as to increase its glass path by $\Delta L$, the time delay incurred is $\frac{\Delta L}{c} [N \! - \! \frac{1}{\cos(\Delta \theta/2)}]$, where $N$ is the relevant glass index and $\Delta \theta$ is the prism  deflection (here, $\Delta \theta \! = \! 56.66^\circ$ at $\omega_d$).  From the primary peak locations in Fig.~3 (corresponding to $\Delta L$ removed from the signal path and $\Delta L$ added to the idler path, in multiples of $3.50000 \, \mathrm{mm}$), the values of $N$ obtained are 1.7287, 1.7282, and 1.7280 in, respectively, cases (b), (c), and (d).  The mean (1.7283) is within 0.0002 of the group index of SF10 glass (1.7281 at $\omega_d$), which is 0.0259 higher than the usual refractive index (1.7022).  Thus Fig.~3 provides a second verification, from an approach different from Ref.~\onlinecite{Stein92}, that photons travel at group velocity.

A classical calculation \cite{Shapiro10}, based on detector cross-correlation, has claimed to predict Franson cancellation.  Elsewhere, this interpretation has been criticized \cite{Franson10}.  In any case, Ref.~\onlinecite{Shapiro10} does not agree with Figs.~2-3 since it  predicts a background that is fully absent in data.  The local dispersion cancellation of the two-photon interferometer \cite{Stein92} \textit{does} have a classical analog \cite{Kaltenbaek08}, and the local nature of the different type of cancellation in Fig.~3 cannot preclude a classical analog here also.  However, an inequality has been derived \cite{Wasak10} showing when the level of nonlocal Franson cancellation is inconsistent with classical theory; it may be possible to develop a similar inequality, applicable to the experimental conditions (local upconversion), to show the cancellation in Fig.~3 to be nonclassical even though it is local.

It is also of interest to compare these observations with Ref.~\onlinecite{Harris07}, which proposes to create a two-photon state with a nonlinear crystal having chirped poling and thus extremely wide bandwidth.  The two-photon state produced is itself chirped, which is proposed to be corrected through multi-order phase correction of only \textit{one} of the modes.  The compressed state is termed a \textit{single-cycle} biphoton since the calculated $R(\tau)$ has width only 1.3 times the optical period at $\omega_d$.  This research remains an experimental challenge, though a possible approach to chirp-correction has been discussed \cite{Brida09}.

The results of Fig.~3 may be regarded as a simplified realization of the proposal of Ref.~\onlinecite{Harris07}; certainly Fig.~3 demonstrates that the GDD of one mode compensates the other.  The mean peak width of Fig.~3 is larger, at 6.8 times the optical period at $\omega_d$; this difference is due to the narrower SPDC bandwidth of the periodically-poled crystal used here.  The present work may thus be considered as the experimental demonstration of \textit{few-cycle} biphotons.  Further, there is subtle evidence in Fig.~3 that this experiment is near its limit of compression, as may be verified by ray-tracing the system of Fig.~1 to obtain  $\phi_{s,i}(\omega)$.  It is found that the $n \! = \! 3$ coefficient $ [ \phi_s^{(3)} \!  - \! \phi_i^{(3)} ] $ of Eq.~(\ref{SPDCPhase}) becomes increasingly negative in the cases of Fig.~3 and, with a stationary-phase approximation in Eq.~(\ref{UCState}), is thus consistent with the increasing right-skewness most apparent in Fig.~3's secondary maxima.  If $R(\tau)$ is evaluated from Eq.~(\ref{UCState}) with a SPDC bandwidth of, say, 1.5 times that of the experiment, this skewness becomes objectionable.  These results are not shown, but other methods \cite{Brida09} may be required to compress larger SPDC bandwidths.

In summary, the dispersion cancellation occurring for frequency-entangled photon pairs, as proposed by Franson  \cite{Franson92,Franson07}, has been investigated.  The approach employs time-resolved upconversion of the pairs, producing fs-level resolution and observed GDD sensitivity of $\approx \! 20 \, \mathrm{fs}^2$.  The cancellation of more than $\pm 10^3 \, \mathrm{fs}^2$ of GDD in the signal $(-)$ and idler $(+)$ modes has been demonstrated.    The observations have direct relevance to research on the generation, compression, and characterization of ultrashort biphotons.

\begin{acknowledgments}
This research was supported by CONACYT (Mexico) under Grant 49570-F.
\end{acknowledgments}

\end{document}